\documentclass[10pt,aps,pra,twocolumn,superscriptaddress]{revtex4-2}
\usepackage{amsfonts, mathtools}
\usepackage{siunitx}
\usepackage{braket}
\usepackage[breaklinks=true,colorlinks,allcolors=blue]{hyperref}
\usepackage{orcidlink}

\newcommand{\trace}[1]{\text{tr}[#1]}

\newcommand{\proj}[1]{\ket{#1}\!\bra{#1}}

\newcommand{\eqlabel}[1]{Eq.~\eqref{#1}}
\newcommand{\figlabel}[1]{Fig.~\ref{#1}}

\begin{document}

\title{Impact of thermal and dissipative effects in a periodically-kicked quantum battery}
\author{Sebastián V. Romero$^{\orcidlink{0000-0002-4675-4452}}$}
\email{sebastian.v.romero@csic.es}
\affiliation{Instituto de Ciencia de Materiales de Madrid ICMM-CSIC, Cantoblanco, 28049 Madrid, Spain}
\affiliation{\mbox{Departamento de Física Teórica de la Materia Condensada, Universidad Autónoma de Madrid, 28049 Madrid, Spain}}

\author{Xi Chen$^{\orcidlink{0000-0003-4221-4288}}$}
\email{xi.chen@csic.es}
\affiliation{Instituto de Ciencia de Materiales de Madrid ICMM-CSIC, Cantoblanco, 28049 Madrid, Spain}

\author{Yue Ban$^{\orcidlink{0000-0003-1764-4470}}$}
\email{yue.ban@csic.es}
\affiliation{Instituto de Ciencia de Materiales de Madrid ICMM-CSIC, Cantoblanco, 28049 Madrid, Spain}
\date{\today}

\begin{abstract}
    Quantum batteries (QBs) have emerged as a promising route for fast energy storage and on-chip power supply in quantum devices. Given the limited analytical understanding of open Floquet QBs, we employ the kicked-Ising model as a tractable platform to systematically study its performance under realistic conditions, including finite temperature effects and environmental dissipation. Starting from Gibbs states of the transverse-field Ising model, we incorporate thermal and decoherence effects along the evolution, using both analytical and numerical approaches. Taking ergotropy as a central figure of merit, we characterize the injected and extractable energy, and identify regimes where charging remains robust despite environmental effects. Our results provide a systematic framework for assessing QB performance under thermal and dissipative effects.
\end{abstract}

\maketitle

The foundational role of Gibbs states as thermal equilibrium descriptors pose them as a natural starting point for quantum batteries (QBs), defined as quantum systems that store and supply energy in the form of work~\cite{alicki2013entanglement, hovhannisyan2013entanglement, binder2015quantacell, campaioli2017enhancing}. Their description relies on concepts from quantum thermodynamics which quantify the maximum extractable work, namely ergotropy, from a quantum state~\cite{pusz1978passive,allahverdyan2004maximal}. Charging protocols exploit coherent driving and collective interactions to generate non-passive states, with Gibbs state initialization and dissipative effects providing a physically-motivated setting to measure ergotropy.%

The quest for scalable QBs has led to a plethora of different proposals, with the kicked-Ising QB standing out as a minimal platform combining robustness against disorder, near-term implementability and analytical tractability~\cite{romero2025kicked}. Described as an Ising chain with a periodically-kicked transverse field, a kicked-Ising chain (KIC) exhibits maximal entanglement growth at the self-dual operator regime~\cite{Akila_2016}, which is leveraged to maximally charge through a stable and resilient protocol. However, existing studies focus exclusively on unitary dynamics, neglecting the unavoidable presence of thermal fluctuations and environmental dissipation in implementations. This raises the question of how robust these charging protocols remain under realistic conditions.

In this work, we present a systematic study of periodically driven QBs under finite temperature and dissipative effects (see~\figlabel{fig:schematic}). To reflect potential environmental couplings and ensure that the battery starts in a passive state~\cite{pusz1978passive}, we initialize the QB with Gibbs states, using the transverse field Ising model (TFIM) as Hamiltonian, to analytically study the dependence with temperature of the injected and extractable energy. Additionally, we test its robustness against non-unitary evolutions including pure dephasing and relaxation-excitation effects, which determine the coherence and relaxation times $T_2$ and $T_1$~\cite{fossfeig2013dynamical}. Our results establish an analytically-supported study when different sources of error are accounted, distancing from the ideal zero-temperature case, enabling to analyze the kicked-Ising QB performance under more realistic conditions.%
\begin{figure}[!tb]
    \centering
    \includegraphics[width=\linewidth]{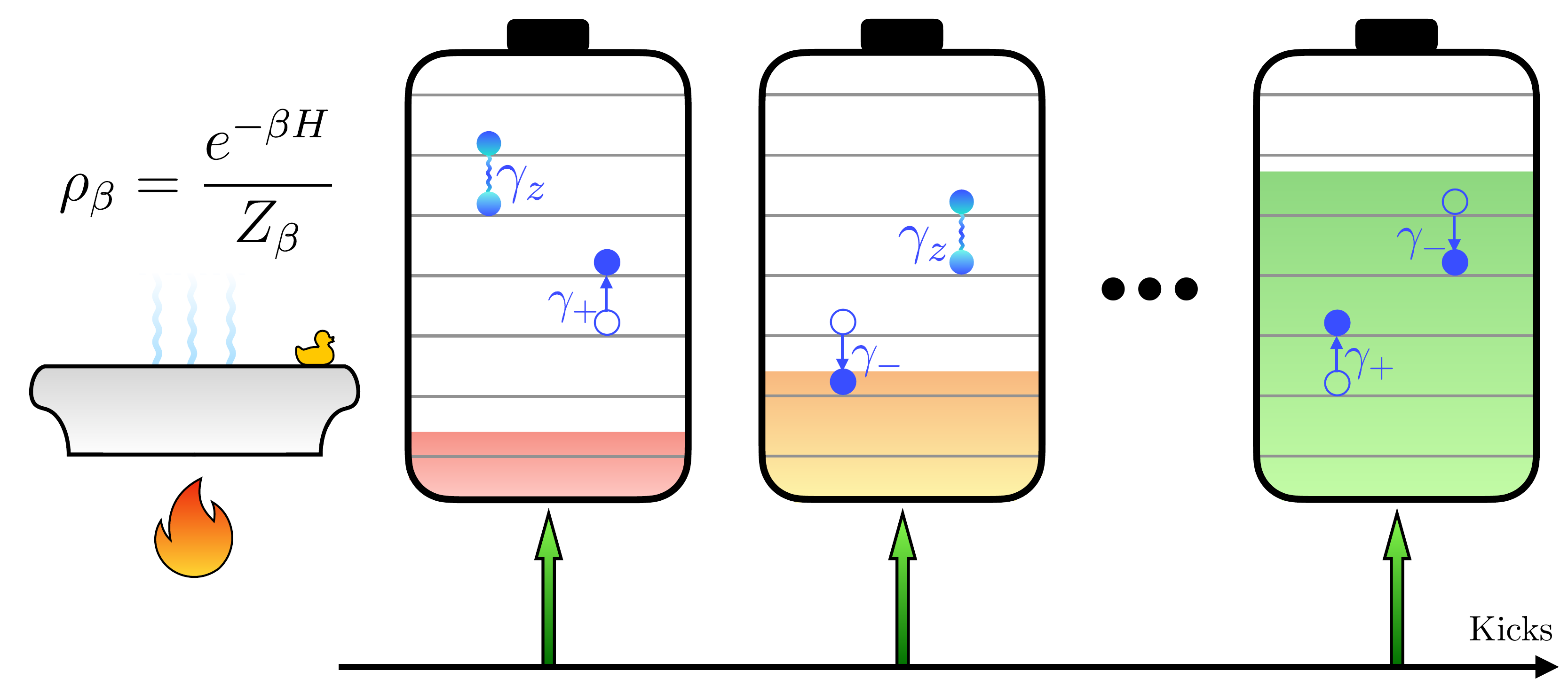}%
    \caption{Schematic of a dissipative kicked-Ising QB. Starting from a passive Gibbs state at temperature $T=1/\beta$, the battery undergoes periodic driving in the presence of dissipation. Decoherence mechanisms include pure dephasing, which suppresses quantum coherences at a rate $\gamma_z$, and excitation-relaxation processes, which induce transitions between energy levels at rates $\gamma_\pm$, respectively.}\label{fig:schematic}\vspace{-3mm}%
\end{figure}%

\emph{Battery setup.}---We use $H(t)=H_0 + \lambda(t)[H_1(t) - H_0]$ as charging protocol, with $H_0 = ({g}/2)\sum_{i=1}^N \sigma^z_i$ the battery Hamiltonian of $N$ quantum cells with eigenvalues $\epsilon_\ell=(\ell-N/2)g$ (multiplicity $\binom{N}{\ell}$) $\forall \ell\in[0,N]$~\cite{romero2024scramblingchargingquantumbatteries}, $H_1$ the charger Hamiltonian and $\lambda(t)$ a function that toggles $H_0$ and $H_1$, where we consider a unit step function $\forall t\in[0,\tau]$ and zero otherwise. Using as initial density matrix $\rho(0)$, the state is evolved as $\rho(\tau) = U(\tau)\rho(0)U^\dagger(\tau)$ with $U(\tau)=\mathfrak{T}\exp[ -i\int_0^\tau \text{d}t H(t)]$ as the unitary time-evolved operator after charging completion, where $\mathfrak{T}\exp[\cdot]$ denotes time-ordering. The mean energy stored is given by $E_N(\tau) = \trace{\rho(\tau)H_0}$.

To quantify the extractable work, we consider the spectral decompositions of the battery Hamiltonian $H_0 = \sum_{j} \varepsilon_j \proj{\varepsilon_j}$ ($\varepsilon_j \le \varepsilon_{j+1}$), and state $\rho = \sum_{j} r_j \proj{r_j}$ ($r_j \ge r_{j+1}$). The ergotropy $\mathcal{W}(\rho,H_0)$ is defined as the maximum work extractable via unitary operations~\cite{allahverdyan2004maximal}, with $\mathcal{W}(\rho,H_0) = \trace{\rho H_0} - \trace{\rho_{\text{p}} H_0}$, where $\rho_{\text{p}} = \sum_{j} r_j \proj{\varepsilon_j}$ is the passive state associated with $\rho$~\cite{pusz1978passive}. By construction, $\rho_{\text{p}}$ has populations that are non-increasing with energy and therefore yields zero ergotropy. Hence, ergotropy is always lower or equal than the energy stored.%

In our setup, we study as charger Hamiltonian a KIC%
\begin{equation}\label{eq:charger}
    H_1(t) = H_I + H_K\sum_{m\in\mathbb{Z}} \delta(t-m),
\end{equation}
considering a unit time interval between kicks. The Ising interaction and the kicked transverse-field terms read as
\begin{equation}\label{eq:components_xx}
    H_I = \sum_{\braket{ij}} J_{ij}\sigma^x_i \sigma^x_j, \qquad H_K = \sum_{i=1}^N b_i\sigma^z_i,
\end{equation}
where $\braket{\cdot}$ denotes nearest-neighbor pairs. The corresponding Floquet operator after one kick reads as $U(1) = \mathfrak{T}\exp[ -i\int_0^1 \text{d}t H(t)] = e^{-iH_K}e^{-iH_I}$, satisfying that $U(m)=U^m(1)$ after $m$ kicks. Hereinafter, we consider periodic boundary conditions ($\sigma^x_{N+1}=\sigma^x_1$), natural units $(\hbar\equiv k_\text{B}\equiv 1$), thus energies and times are in units of $|J|$ and $1/|J|$, and we set ${g}=1$. The $J_{ij}$ couplings set the strength among spins $(i,j)$ and $b_i$ the strength of the transverse field applied on spin $i$, working at the self-dual point $J_{ij}=J=\pi/4$, $b_i=b=-\pi/4$~\cite{Akila_2016}.

We initialize our QB in a Gibbs state of the form%
\begin{equation}\label{eq:initial}
    \rho(0) = \frac{e^{-\beta H_\text{th}}}{Z}, \qquad \text{with }Z=\trace{e^{-\beta H_\text{th}}}
\end{equation}
and $\beta=1/T$ the inverse temperature, $H_\text{th}$ the Gibbs state Hamiltonian and $Z$ the partition function. Two limiting cases are of particular interest. In the infinite-temperature limit, $\lim_{\beta\to 0} \rho(0) = 1/2^N$ corresponds to the completely mixed state. In contrast, in the zero-temperature limit, $\lim_{\beta\to\infty} \rho(0) = \proj{\text{GS}}$, i.e., the system projects onto the ground state of $H_{\text{th}}$. This follows from the interpretation of the Gibbs state as an imaginary-time propagator (with $\tau=\beta$), whose long-time behavior is dominated by the lowest-energy eigenstate~\cite{simon2005functional}. Remarkably, states of the form of~\eqlabel{eq:initial} are completely passive~\cite{pusz1978passive}, thus no work can be extracted from them, motivating its inclusion in our studies.

Finally, the thermal expectation value of an observable $O$ can be decomposed into parity sectors as $\braket{O} = (Z_+/Z)\braket{O}_+ + (Z_-/Z)\braket{O}_-$, where $P = \prod_{i=1}^N \sigma_i^z$ is the parity operator, $P_\pm = (1 \pm P)/2$ are the projectors onto the even $(+)$ and odd $(-)$ subspaces, and $\braket{\cdot}_\pm = \trace{e^{-\beta H^\pm_\text{th}}\cdot}/Z_\pm$ with $Z_\pm = \trace{e^{-\beta H^\pm_\text{th}}}$ and  $H_{\text{th}}^\pm$ the restriction of $H_{\text{th}}$ to the corresponding parity sector.

\emph{TFIM thermal initialization.}---We consider the TFIM Hamiltonian $H_\text{th} = J_\text{th}\sum_{\braket{ij}} \sigma^x_i\sigma^x_j + h_\text{th}\sum_{i=1}^N \sigma^z_i$ to initialize the state [\eqlabel{eq:initial}]. This model can be exactly diagonalized via the Jordan-Wigner and Fourier transformation~\cite{sm}, yielding%
\begin{equation}
\begin{split}
    H_\text{th} &= \sum_k \left[ \epsilon_k c^\dagger_kc_k -\frac{\Delta_k}{2} (c^\dagger_kc^\dagger_{-k} + c_{-k}c_k) \right] + \text{const.} \\
    &= \sum_k \Psi^\dagger_k \mathcal{H}_{\text{th},k}\Psi_k + \text{const.},
\end{split}
\end{equation}%
where $\epsilon_k = 2(J_{\text{th}} \cos k - h_{\text{th}})$ and $\Delta_k = 2J_{\text{th}} \sin k$. 
Here, $\mathcal{H}_{\text{th},k} = \epsilon_k \tau^z - \Delta_k \tau^x$ is the Bogoliubov-de Gennes Hamiltonian in the Nambu basis $\Psi_k = [c_k, c_{-k}^\dagger]^{\text{T}}$, with eigenvalues $\lambda^\pm_k = \pm E_k = \pm \sqrt{\epsilon_k^2 + \Delta_k^2}$ and Pauli matrices $\tau^{x,y,z}$. Note that the same strategy can be used to diagonalize~\eqlabel{eq:charger}~\cite{romero2025kicked, sm}. The Hamiltonian $H_\text{th}$ is split into $(k,-k)$ pairs with momentum grids $k\in K_\pm = \{(2n+p)\pi/N \,|\, n=0,\dots,N-1\}$ for the even ($p=1$) and odd ($p=0$) parity sectors, respectively. Each $(k,-k)$ sector spans a four-dimensional Hilbert space, $\text{span}\{\ket{0_k 0_{-k}}, \ket{1_k 1_{-k}}, \ket{1_k 0_{-k}}, \ket{0_k 1_{-k}}\}$. While the even sector encapsulates the non-trivial dynamics, the odd sector presents a static contribution.%
\begin{figure*}[!t]
  \centering
  \includegraphics[width=\linewidth]{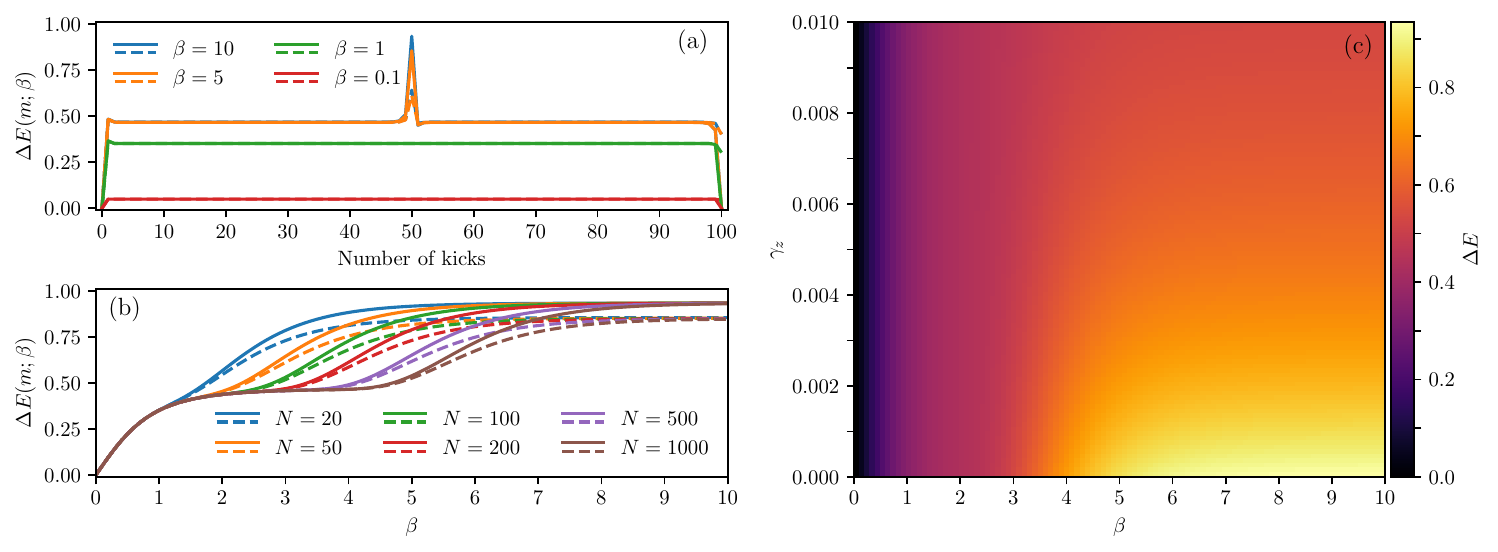}\vspace{-2mm}%
  \caption{Normalized injected energies using $(J_\text{th},h_\text{th})=(1/2,1)$ for the initial TFIM and evolving at the self-dual point. Dashed (solid) lines correspond to dynamics in the presence (absence) of dephasing. (a) For $N = 100$ spins, evolution for different $\beta$ up to $m = 100$ kicks with $\gamma_z = 0.005$. (b) Normalized injected energies at $m = N/2$ for different system sizes across varying temperatures, with $\gamma_z = 1/(10N)$. (c) Normalized injected energy in the presence of dephasing under different $(\beta, \gamma_z)$ pairs. All data are obtained analytically~\cite{sm}.}\label{fig:qb1}
\end{figure*}

Therefore, the time evolution per $k$-mode under~\eqlabel{eq:charger} in the even sector is described by $\rho^+_k(m)= U_k(m)\rho^+_k(0)U^\dagger_k(m)$ with initial state
\begin{equation}
    \rho^+_k(0)= \frac{e^{-\beta H^+_{\text{th},k}}}{\trace{e^{-\beta H^+_{\text{th},k}}}} = \frac{1}{2}+\frac{\tanh\beta E_k}{2E_k}(\epsilon_k\tau^z + \Delta_k\tau^x).
\end{equation}
The single-period Floquet operator reads as $U_k(1)= \mathfrak{T}\exp[-i\int^1_0 \text{d}t H_k(t)]$ with $U_k(m)=U^m_k(1)$ and $H_k(t)= 2J\cos k\tau^z - 2b\sin k\sum_{m\in\mathbb{Z}}\delta(t-m)\tau^x$. After applying the Jordan-Wigner transformation and Fourier transform to $H_0 = \sum_k (1-2c^\dagger_kc_k) = \sum_k O_k$, the energy stored after $m$ kicks can be computed using that~\cite{sm}%
\begin{widetext}\vspace{-2mm}
\begin{equation}\label{eq:obs}
\begin{split}
    \braket{O^+_k(m)} &= \trace{O^+_k\rho^+_k(m)} = \frac{2\tanh \beta E_k}{E_k} \left[ \epsilon_k(1-2|\beta_k(m)|^2) + 2\Delta_k\text{Re}[\alpha_k(m)\beta_k(m)] \right], \quad 
\end{split}
\text{with }U_k(m)\!=\!\begin{bmatrix}
        \alpha_k(m) & -\beta^*_k(m) \\ \beta_k(m) & \alpha^*_k(m) \end{bmatrix}
\end{equation}\vspace{-2.5mm}
\end{widetext}%
and averaging over parity sectors, using that the parity operator $P$ commutes both with $H(t)$ and $H_\text{th}$, thus parity remains constant along the evolution, and that it decomposes as $P=\prod_k P_k$. In the local limit ($J_\text{th}=0$) and defining $m^z_\beta= \tanh\beta h_\text{th}$, we obtain instead that 
\begin{equation}
    \braket{O^+_k(m)} = {g}m^z_\beta \bigg[ \sin^22J \sin^2k\frac{\sin^2 m\theta_k}{\sin^2\theta_k} - \frac{1}{2} \bigg],
\end{equation}
with $\theta_k=\cos^{-1}\text{Re}[\alpha_k(1)]$, which reduces at the self-dual point to $\braket{O^+_k(m)} = {g}m^z_\beta[ \sin^2mk - 1/2 ]$. Summing over modes yields the next closed-form results. For even $N$,
\begin{equation}
    E_N(m;\beta) = \frac{gN}{2}\cdot\begin{cases}
        -m^z_\beta & \!\text{if }m\equiv 0 \!\!\!\pmod{N} \\
        (m^z_\beta)^{N-1} & \!\text{if }m\equiv \frac{N}{2} \!\!\!\pmod{N} \\
        0 & \!\text{otherwise}
    \end{cases},
\end{equation}
while for odd $N$ we obtain
\begin{equation}
    E_N(m;\beta) = \frac{gN}{2}\cdot\begin{cases}
        -m^z_\beta & \!\text{if }m\equiv 0 \!\!\!\pmod{N} \\
        0 & \!\text{otherwise}
    \end{cases}.
\end{equation}
Defining the normalized energy injected as $\Delta E_N(m;\beta) = (E_N(m;\beta) - E_N(0;\beta))/N$, one finds that for even $N$, it is maximized at $m=N/2$, yielding $\Delta E_N(N/2;\beta) = ({g}/2) [m^z_\beta + (m^z_\beta)^{N-1}]$. This expression interpolates between zero and infinite temperature limits $\lim_{\beta\to\{0,\infty\}} \Delta E_N(N/2;\beta) = \{0,\text{sgn}(h_\text{th}){g}\}$, and, in the thermodynamic limit $\lim_{N\to\infty}\Delta E_N(N/2;\beta)=({g}/2)m^z_\beta$.%

\emph{Open QB: ergotropy analysis.}---Now we consider an open-system framework where the dynamics is governed by the Lindblad master equation
\begin{equation}
  \frac{\text{d}\rho(t)}{\text{d}t} = -i[H(t), \rho(t)] + \sum_i \! \left[ L_i\rho(t)L^\dagger_i - \frac{1}{2}\left\{L^\dagger_iL_i,\rho(t)\right\} \right]
\end{equation}
with the Lindblad operators $L_i$ describing the coupling to the environment. In our studies we consider dephasing with a damping rate $\gamma_z$, namely $L_i=\sqrt{\gamma_z}\sigma^z_i$,
leading to%
\begin{equation}\label{eq:dephasing}
  \frac{\text{d}\rho(t)}{\text{d}t} = -i[H(t), \rho(t)] + \gamma_z\sum_{i=1}^N \left( \sigma^z_i\rho(t)\sigma^z_i - \rho(t) \right).
\end{equation} 
This form captures pure dephasing processes that suppress quantum coherences, directly related to $T_2$ coherence times as $T_2=T_\phi = 1/(2\gamma_z)$~\cite{sm}, where $T_\phi$ quantifies the timescale over which quantum superpositions decay, thereby affecting the ergotropy. Notably, the energy can be computed analytically: after a Jordan-Wigner transformation followed by a Fourier transform, Eq.~\eqref{eq:dephasing} reduces to a system of $3N/2$ coupled linear ordinary differential equations, which can be solved in linear time~\cite{sm}.

In~\figlabel{fig:qb1} we compute the normalized injected energies for different temperatures $\beta\in[0,10]$, even system sizes ranging from $N=10$ to $N=1000$ and decoherence rates $\gamma_z\in[0,0.01]$, initializing the state with $(J_\text{th},h_\text{th})=(1/2,1)$. While for lower temperatures the protocol can inject energy peaking at $m=N/2$, as expected~\cite{romero2025kicked}, for $\beta\to 0$ the state converges into a complete passive state, where no energy is stored. Moreover, with increasing system size and $\gamma_z=0$, the normalized energy injected saturates to the thermodynamic limit of~\eqlabel{eq:obs}, which is restored by substituting $\sum_k \mapsto (N/2\pi)\int_{-\pi}^\pi \text{d}k$. In~\figlabel{fig:qb1}c, for a KIC QB of $N=100$ spins at the self-dual point, we can see the impact of the initial temperature and the decoherence rate in the energy injected at $m=N/2$. As expected, we observe that for increasing $\beta$, thus we approach the zero temperature limit, the injected energy increases. Regarding decoherence, we observe that for increasing $\gamma_z$ and $\beta$ the normalized injected energy saturates to $1/2$, which results from the initial state converging to a classical incoherent steady state $\rho_\text{ss}=1/2^N$, so a completely mixed state. In particular, coherences decay exponentially with $\gamma_z m$. Therefore, when $\gamma_z>0$, one can check the robustness against decoherence by comparing this product with respect the KIC parameters, which set the energy scaling. In other words, when $\gamma_z m \ll J,b$ the coherent dynamics dominate, while for $\gamma_z m \gg J,b$ coherences are heavily suppressed.%
\begin{figure*}[!tb]
    \centering
    \includegraphics[width=\linewidth]{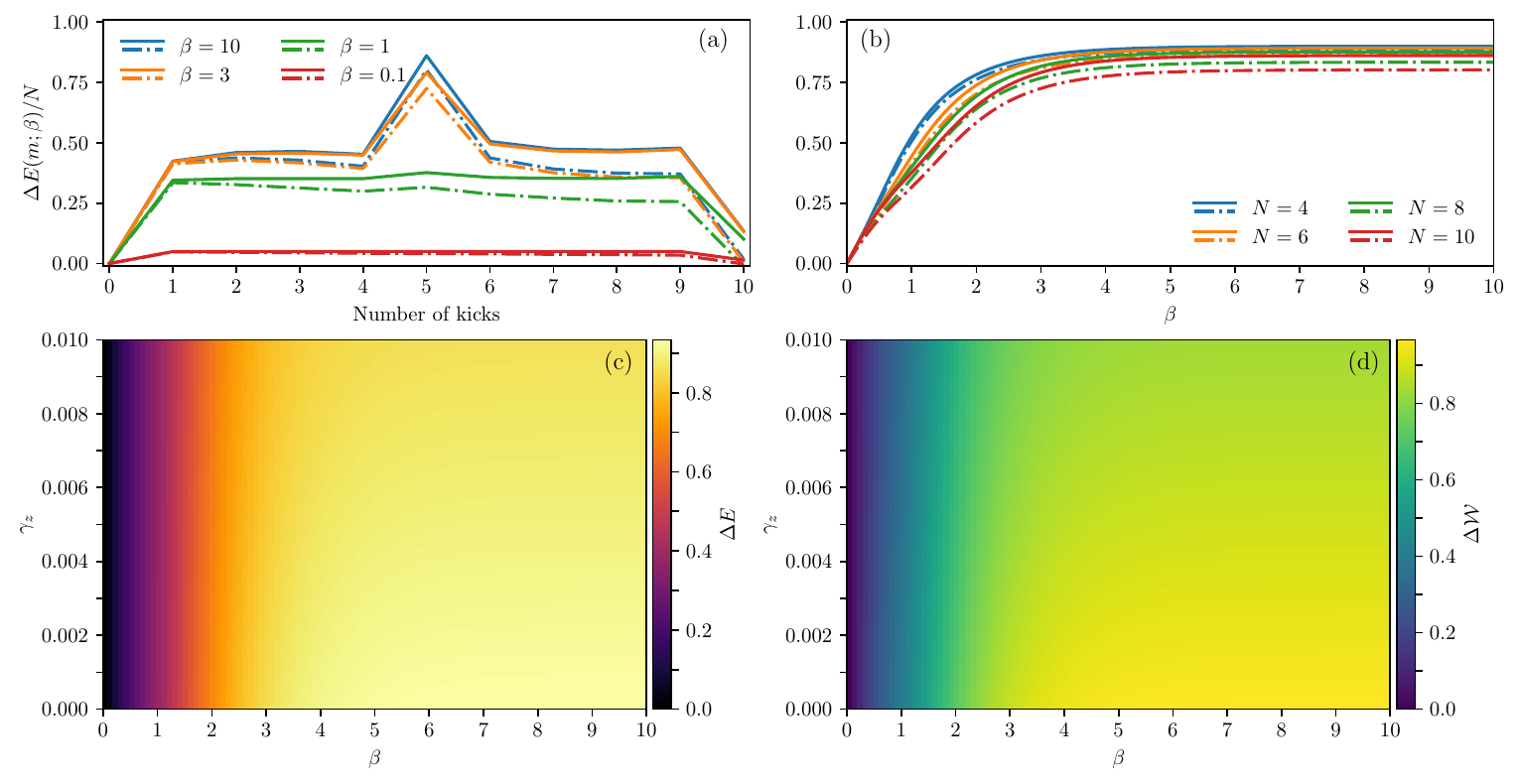}%
    \caption{(a) Normalized injected energies (solid) and ergotropies (dash-dotted) in the presence of dephasing using $(J_\text{th},h_\text{th})=(1/2,1)$ for the initial TFIM Hamiltonian and evolving at the self-dual point $(J,b)=(\pi/4,-\pi/4)$ using $N=10$ spins up to $m=10$ kicks with $\gamma_z=0.01$. (b) Same analysis as in (a), but for different system sizes $N$ and temperatures $\beta$ at kick $m=N/2$. (c) Normalized injected energy at the kick $m=N/2$ of a KIC QB with $N=10$ in the presence of dephasing under various $(\beta,\gamma_z)$ pairs. We set $(J_\text{th},h_\text{th}) = (1/2,1)$ for the TFIM Hamiltonian. (d) Same study as (c), but computing the ergotropy.}\label{fig:decoh}
\end{figure*}

To compare how much work can be extracted from the injected energy under decoherence, we first define the normalized ergotropy $\Delta\mathcal{W}(m;\beta)= (\mathcal{W}(m;\beta) - \mathcal{W}(0;\beta))/N$ and numerically compute both quantities up to $N=10$. In~\figlabel{fig:decoh}, we plot them for a KIC QB at the self-dual point, studying the impact of the initial state and decoherence rates present in the system, where we obtain similar plots as in~\figlabel{fig:qb1}, with ergotropy decaying with increasing number of kicks applied.%

In addition, we also consider excitation-relaxation processes with rates $\gamma_\pm$, respectively, which can be described by the Lindblad master equation
\begin{equation}\label{eq:ladder}
\begin{split}
  \frac{\text{d}\rho(t)}{\text{d}t} &= -i[H(t), \rho(t)] \\
  &+ \sum_{s\in\pm}\gamma_s\sum_{i=1}^N \left[ \sigma^s_i\rho(t)\sigma^{s\dagger}_i - \frac{1}{2}\left\{ \sigma^{s\dagger}_i\sigma^s_i, \rho(t)\right\} \right],
\end{split}
\end{equation}
with $\sigma^\pm_i= (\sigma^x_i \pm i\sigma^y_i)/2$ the ladder operators. These Lindblad operators encode external factors that promote states to higher or lower excited states at a rate $\gamma_\pm$.%

In particular, including spontaneous emission and absorption contributions is essential for modeling quantum systems coupled to finite-temperature environments, as these processes govern the approach to thermal equilibrium involving energy exchanges, as for quantum optics setups. To account these effects, \eqlabel{eq:ladder} can be considered with emission and absorption rates $\gamma_- = \gamma ( n_\text{th} + 1 )$ and $\gamma_+ = \gamma n_\text{th}$, where $n_\text{th} = 1/(e^{\beta\omega_0}-1)$ is the Bose-Einstein occupation number , $\beta$ is the inverse temperature of the bath and $\omega_0$ the frequency of the bath mode, setting $\omega_0=1$. These processes are related to both $T_1$ and $T_2$ coherence times as $1/T_1=\gamma_+ + \gamma_-$ and $T_2 = 2T_1$~\cite{sm}, where $T_1$ characterizes the relaxation of excited-state populations and thus directly impacts the battery performance. The dissipative rates satisfy that $ \gamma_+/\gamma_- = e^{-\beta\omega_0 } $, condition that the ratio of populations meet at equilibrium. Contrarily to the previous case, now numerical methods are needed since~\eqlabel{eq:ladder} contains higher-order terms after Jordan-Wigner transformation, disabling its exact diagonalization.%
\begin{figure*}[!tb]
    \centering
    \includegraphics[width=\linewidth]{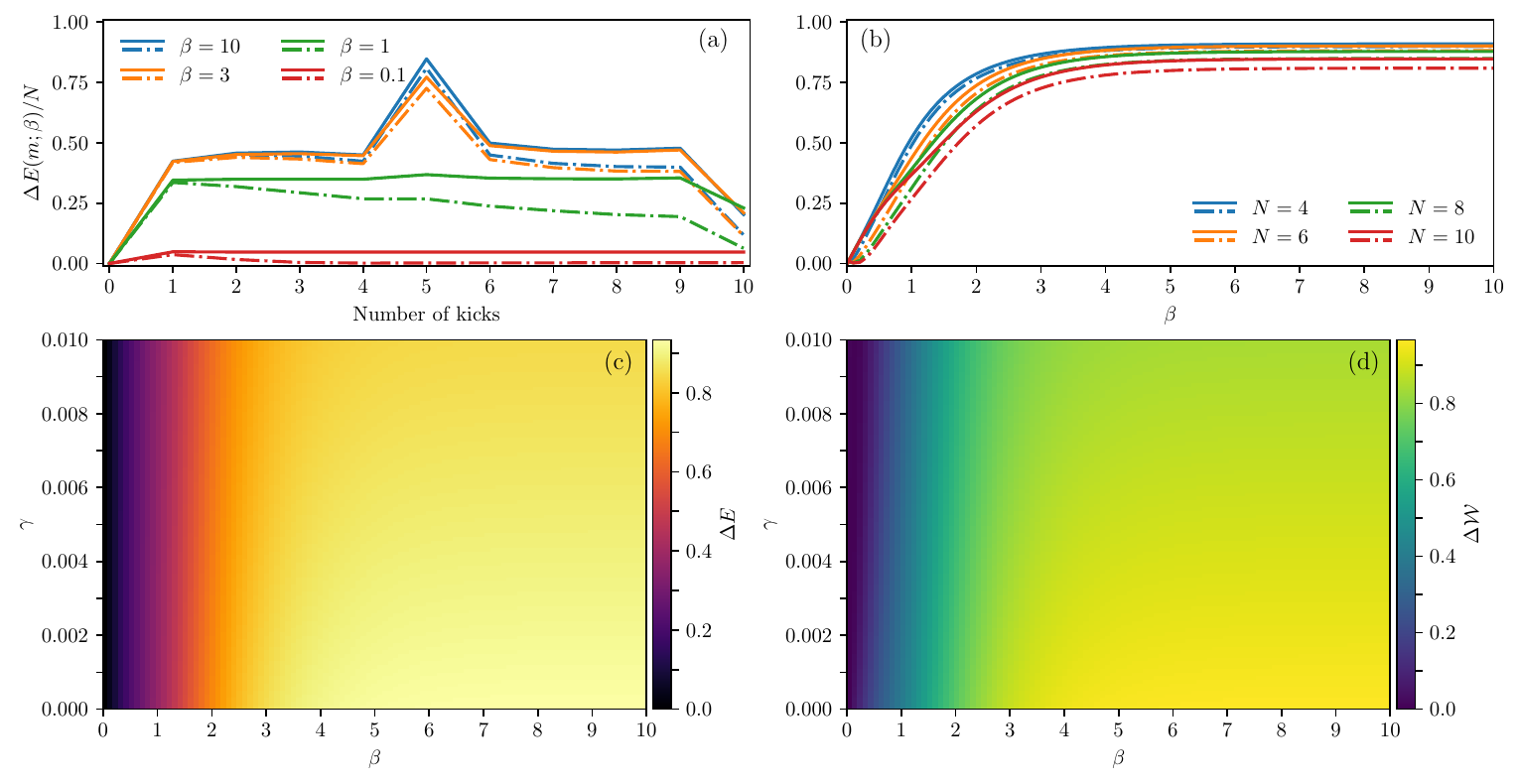}%
    \caption{(a) Normalized injected energies (solid) and ergotropies (dash-dotted) in the presence of thermal dissipation using $(J_\text{th},h_\text{th})=(1/2,1)$ for the initial TFIM Hamiltonian and evolving at the self-dual point $(J,b)=(\pi/4,-\pi/4)$. Results are shown for $N=10$ spins, up to $m=10$ kicks with $\gamma=0.01$. (b) Same analysis as (a) but for different system sizes $N$ across different temperatures $\beta$ at the kick $m=N/2$. (c) Normalized injected energy at the kick $m=N/2$ of a KIC QB with $N=10$ in the presence of thermal dissipation under various $(\beta,\gamma)$ pairs. We set $(J_\text{th},h_\text{th}) = (1/2,1)$ for the TFIM Hamiltonian. (d) Same study as (c), but computing the ergotropy.}\label{fig:diss}
\end{figure*}

In~\figlabel{fig:diss} we compute both the normalized injected energy and ergotropy for a KIC QB of $N=10$ spins at the self-dual point under thermal dissipation after applying $m=N/2$ kicks, studying again the impact of the temperature set and the rate of dissipation present in the system considering $(\beta,\gamma)\in[0,10]\times[0,0.01]$. We observe similar plots as for~\figlabel{fig:decoh}, where the lower the temperature and lower values of $\gamma$, the higher injected energies and ergotropies obtained, as expected. Moreover, we can observe that our charging protocol is robust against decoherence. In current platforms, environmental noise is typically dominated by decoherence processes, including both energy relaxation and dephasing, with the latter often providing the leading contribution to coherence loss in neutral atoms and trapped ions~\cite{saffman2005analysis, wineland1998experimental} and remaining a major limitation also in solid-state qubits~\cite{chirolli2008decoherence}.%

\emph{Towards an experimental realization.}---Among other candidates, ultracold atoms in optical lattices or Rydberg arrays~\cite{johnson2008rabi,beguin2013direct,hermann2014long,labuhn2016tunable,guardado-sanchez2018probing,graham2019rydberg} feature MHz-scale Ising interactions and Rabi frequencies, with coherence times reaching hundreds of microseconds at temperatures ranging from micro to nanokelvins. Notably, long-range interactions do not affect charging performance remarkably~\cite{romero2025kicked}. For typical van der Waals strengths $C_6\simeq\SI{10}{GHz\,\micro\meter^6}$, blockade radii $R_b\simeq\SI{5}{\micro\meter}$, Rabi frequencies $\Omega\simeq\SI{1}{MHz}$ and coherence times $T_2^*\simeq\SI{100}{\micro\second}$ one obtains an effective coherence-decay rate of order $\gamma_{\text{phys}} \sim 1/(2T_2^*) \simeq 5\,\text{kHz}$. This suggests that on the order of $100$ kicks could be implemented with a kick period of $\SI{1}{\micro\second/kick}$ while remaining within the available coherence window. Moreover, taking as a representative temperature $T_\text{phys}\simeq\SI{1}{\nano\kelvin}$ and $\beta=1/T=10$ as target dimensionless temperature, at thermal equilibrium the effective energy scale is given by $J_\text{th} = k_\text{B} T_\text{phys} / (\hbar T) \simeq \SI{1}{kHz}$. %

Trapped ions are also a natural setup~\cite{kim2010quantum,britton2012engineered,jurcevic2017direct,zhang2017observation}, with current setups characterized by $J_{ij}$ and $b_i$ in~\eqlabel{eq:components_xx} at the kHz scale, $T^*_2$ up to seconds and temperatures ranging from micro to milikelvins. Taking $J_{ij},b_i\simeq\SI{10}{kHz}$ and $T^*_2\simeq\SI{100}{ms}$ (thus an effective coherence-decay rate $\gamma_\text{phys}\simeq\SI{5}{Hz}$), we estimate as rate $\SI{0.1}{ms/kick}$ to apply $1000$ kicks within the coherence time. For $T_\text{phys}\simeq \SI{10}{\micro\kelvin}$ and $\beta=10$, we get at thermal equilibrium $J_\text{th} \simeq \SI{10}{MHz}$ as effective energy scale.%

\emph{Conclusion.}---In this work, we have investigated the impact of finite temperature and dissipation on the performance of a periodically-kicked QB. Starting from passive Gibbs states, we have derived analytical expressions for the energy injected, revealing how temperature and system size limit the maximum extractable work. We have further examined the role of dephasing and excitation-relaxation processes, showing how environmental couplings degrade performance while also revealing regimes of robustness. Overall, our results provide insight into the interplay between coherent control and dissipation in QB, contributing to ongoing efforts toward efficient and robust quantum energy storage. These results motivate the exploration of optimal control strategies to mitigate dissipation and improve performance in  realistic implementations.%

\emph{Acknowledgments.}---We thank Alan C. Santos for fruitful discussions. This work is supported by the project grants PID2024-157842OA-I00 and PID2021-126273NB-I00 funded by MCIN/AEI/10.13039/501100011033 and by ``ERDF A way of making Europe'' and ``ERDF Invest in your Future'', Spanish national project in the field of Artificial Intelligence (AIA2025-163435-C44), the Basque Government through Grant No. IT1470-22, the Severo Ochoa Centres of Excellence program through Grant CEX2024-001445-S, the Spanish Ministry of Economic Affairs and Digital Transformation through the QUANTUM ENIA project call-Quantum Spain project.%

\emph{Data availability.}---The data that support the findings of this article are openly available~\cite{repository}.

\bibliography{bibfile}

\end{document}


\title{Supplemental Material for: \texorpdfstring{\\}{}``Impact of thermal and dissipative effects in a periodically-kicked quantum battery''}
\author{Sebastián V. Romero$^{\orcidlink{0000-0002-4675-4452} }$}
\email{sebastian.v.romero@csic.es}
\affiliation{Instituto de Ciencia de Materiales de Madrid ICMM-CSIC, Cantoblanco, 28049 Madrid, Spain}
\affiliation{\mbox{Departamento de Física Teórica de la Materia Condensada, Universidad Autónoma de Madrid, 28049 Madrid, Spain}}

\author{Xi Chen$^{\orcidlink{0000-0003-4221-4288}}$}
\email{xi.chen@csic.es}
\affiliation{Instituto de Ciencia de Materiales de Madrid ICMM-CSIC, Cantoblanco, 28049 Madrid, Spain}

\author{Yue Ban$^{\orcidlink{0000-0003-1764-4470}}$}
\email{yue.ban@csic.es}
\affiliation{Instituto de Ciencia de Materiales de Madrid ICMM-CSIC, Cantoblanco, 28049 Madrid, Spain}
\date{\today}

\begin{abstract}
    In this Supplemental Material, we provide a step-by-step derivation of the analytical machinery used to obtain the results and support the findings in the main text. Starting from the exact diagonalization of the kicked-Ising model, we develop an analytical method to include dephasing along the charging process, which can be solved as a coupled linear system of differential equations with the number of variables scaling linearly with system size. Finally, extended notes are provided to account Gibbs thermal states as initial states to compute analytically thermal averages of the energy injected under a periodically-kicked open evolution.
\end{abstract}

\maketitle

\tableofcontents

\section{Exact diagonalization of the kicked-Ising model}

In this section, we obtain analytically the Floquet operator of a periodically-kicked Ising chain and its corresponding powers, essential parts for constructing the dynamics of the kicked-Ising chain quantum battery (QB). A more thorough derivation can be found in Ref.~\cite{romero2025kicked}.

We consider as charger Hamiltonian $H_1(t)=H_I+H_K(t)$ [Eq.~(2) in the main text], with Ising and kicked-transverse field contributions
\begin{equation}\label{eq:ref_ham1}
    H_I = J\sum_{\braket{ij}}\sigma^x_i\sigma^x_j,\qquad H_K(t) = \underbrace{b\sum_{j=1}^N\sigma^z_j}_{= H_K}\sum_{t_i\in\mathbb{Z}}\delta(t-t_i).
\end{equation}
with battery Hamiltonian $H_0=({g}/2)\sum_{i=1}^N\sigma^z_i$. Additionally, we define for convenience $b(t)= b\sum_{t_i\in\mathbb{Z}}\delta(t-t_i)$, which encapsulates the periodically-kicked transverse field coupling strength. 

To begin with the exact diagonalization of this model, we apply the Jordan-Wigner transformation $\sigma^x_i = -(c^\dagger_i + c_i)\prod_{j<i} (1-2c^\dagger_j c_j)$ and $\sigma^z_i = 1-2c^\dagger_i c_i$, with $c^\dagger_i$ ($c_i$) a spinless creation (annihilation) fermion at the $i$th site, following the anticommutation rules $\{c_i,c^\dagger_j\}=\delta_{ij}$ and $\{c_i,c_j\}=0$. Under the spinless fermion representation, the charger Hamiltonian reads as
\begin{equation}\label{eq:reduced_hamiltonian}
    H(t) = J\sum_{\braket{ij}} (c^\dagger_ic_j -c_ic_j +\hc) +b(t)\sum_{i=1}^N (1-2c^\dagger_ic_i),
\end{equation}
This Hamiltonian can be split into its even and odd parity sectors, $H^+$ and $H^-$ respectively, as $H=P^+H^+P^+ + P^-H^-P^-$, where $P^\pm= (1\pm P)/2$ are the projectors on the even and odd subspaces with $P= \prod_{i=1}^N \sigma^z_i$. Here, $H^\pm$ are given by~\eqlabel{eq:reduced_hamiltonian} following antiperiodic (periodic) boundary conditions $c_{N+1}=-c_1$ ($c_{N+1}=c_1$) for the even (odd) parity sector. 

Next step is to apply the Fourier transform 
\begin{equation}\label{eq:fourier}
    c_l=\frac{e^{-i\pi/4}}{\sqrt{N}}\sum_k c_ke^{ik(la)},
\end{equation}
with $a$ the lattice spacing and taking as pseudomomenta $k\in K_\pm = \{(2n+p)\pi/N \,|\, n=0,\dots,N-1\}$ for the even ($p=1$) and odd ($p=0$) parity sectors, respectively, which has to be consistent with the periodic boundary conditions chosen in the Jordan-Wigner transformation. The total Hilbert space is now a product of two-dimensional spaces spanned by the states $\ket{0_k0_{-k}}$ and $\ket{1_k1_{-k}}$ for each $k$ mode. After this transform, the Hamiltonian reads as
\begin{equation}\label{eq:kic_momentum}
    H(t) = \sum_k \left\{ 2[J\cos(ka) - b(t)] c^\dagger_kc_k - J\sin(ka)(c^\dagger_kc^\dagger_{-k} + c_{-k}c_k) + b(t) \right\}.
\end{equation}

These equations can be written in a more convenient form by defining for the $k$th mode
\begin{equation}\label{eq:matricial_bdg}
    H_k(t)= 2[J\cos(ka)-b(t)]\tau^z -2J\sin(ka)\tau^x= 2\begin{bmatrix}
        J\cos(ka)-b(t) & -J\sin(ka) \\
        -J\sin(ka) & b(t)-J\cos(ka)
    \end{bmatrix},
\end{equation}
with $\tau^{\{x,y,z\}}$ the Pauli matrices, satisfying the time-dependent Schr\"odinger-like equation $i\text{d}U_k(t)/\text{d}t=H_k(t)U_k(t)$. The Floquet operator after one kick is given by
\begin{equation}\label{eq:uk_floquet}
\begin{split}
    U_k(1) &= \mathfrak{T}\exp\left[ -i\int_{0}^{1} \text{d}tH_k(t) \right] = \exp(2ib\tau^z)\exp(-2iJ[\cos(ka)\tau^z-\sin(ka)\tau^x]) \\
    &= \begin{bmatrix}
    e^{2ib}[\cos(2J) - i\sin(2J)\cos(ka)] & ie^{2ib}\sin(2J)\sin(ka) \\
    ie^{-2ib}\sin(2J)\sin(ka) & e^{-2ib}[\cos(2J) + i\sin(2J)\cos(ka)]
    \end{bmatrix}=\begin{bmatrix}
        \alpha_k & -\beta^*_k \\
        \beta_k & \alpha^*_k
    \end{bmatrix}.
\end{split}
\end{equation}
satisfying that $U_k(m)=U^m_k(1)$. Relying on the Cayley-Hamilton theorem, where every square matrix satisfies its own characteristic equation, we can compute the $m$th power of the Floquet operator by means of the Chebyshev polynomials of the second kind of degree $m$ in $x$, $\mathcal{U}_m(x)$, as
\begin{equation}\label{eq:power_floquet}
    U^m_k(1) = U_k(1)\mathcal{U}_{m-1}(\xi_k) - \mathcal{U}_{m-2}(\xi_k) = 
    \begin{bmatrix}
        \alpha_k\mathcal{U}_{m-1}(\xi_k)-\mathcal{U}_{m-2}(\xi_k) & -\beta_k^*\mathcal{U}_{m-1}(\xi_k) \\
        \beta_k\mathcal{U}_{m-1}(\xi_k) & \alpha^*_k\mathcal{U}_{m-1}(\xi_k)-\mathcal{U}_{m-2}(\xi_k)
    \end{bmatrix} = 
    \begin{bmatrix}
        \alpha_k(m) & -\beta_k^*(m) \\
        \beta_k(m) & \alpha^*_k(m)
    \end{bmatrix},
\end{equation}
with $\xi_k=\trace{U_k(1)}/2=\text{Re}[\alpha_k]$. It can be further simplified by considering that $\mathcal{U}_m(\cos\theta)=\sin(m+1)\theta/\sin\theta$ with $\theta_k=\cos^{-1}\xi_k = \cos^{-1}[\cos(2b)\cos(2J) + \sin(2b)\sin(2J)\cos(ka)]$, where
\begin{equation}
\begin{split}
  \alpha_k(m) &= \alpha_k\mathcal{U}_{m-1}(\xi_k)-\mathcal{U}_{m-2}(\xi_k) = e^{-2ib}[\cos(2J) + i\sin(2J)\cos(ka)]\frac{\sin m\theta_k}{\sin\theta_k} - \frac{\sin (m-1)\theta_k}{\sin\theta_k}, \\
  \beta_k(m) &= \beta_k\mathcal{U}_{m-1}(\xi_k) = ie^{-2ib}\sin(2J)\sin(ka)\frac{\sin m\theta_k}{\sin\theta_k}.
\end{split}
\end{equation}

As a side note, thermodynamic limit can be restored by substituting $\sum_k \mapsto (N/2\pi)\int^{\pi}_{-\pi}\text{d}k$.

\section{Analytical derivation of the open charging dynamics}

\subsection{Preliminaries: exact solution of the coherent evolution}

Throughout our discussion, we analyze the QB charging performance when the system is open, whose evolution is governed by the Lindblad master equation
\begin{equation}\label{eq:open}
  \dot{\rho}(t) = -i[H(t), \rho(t)] + \sum_i \bigg( L_i\rho(t)L^\dagger_i - \frac{1}{2}\{L^\dagger_iL_i,\rho(t)\} \bigg),
\end{equation}
where the first term of the right hand side encodes the coherent part of the evolution and the latter the dissipative one, with $L_i$ a set of Lindblad operators encapsulating the open part of the evolution.

Since we are interested in the energy injection dynamics, it is more convenient to write the Lindblad master equation in its Heisenberg form. For any observable $O$, its expectation value is given by $\braket{O}(t)=\trace{O\rho(t)}$. In the Schrödinger picture, observables are taken as time-independent obtaining that
\begin{equation}
    \frac{\text{d} \braket{O}(t)}{\text{d}t} = \trace{O\frac{\text{d}\rho(t)}{\text{d}t}} = -i\trace{O[H(t),\rho(t)]} + \sum_i \trace{ O \bigg( L_i\rho(t)L^\dagger_i - \frac{1}{2}\left\{L^\dagger_iL_i,\rho(t)\right\} \bigg) }.
\end{equation}
Using the cyclic property of the trace, we get that
\begin{align}
    \trace{O[H(t),\rho(t)]} &= -\trace{\rho(t)[H(t),O]}, \\
    \trace{ O \bigg( L_i\rho(t)L^\dagger_i - \frac{1}{2}\{L^\dagger_iL_i,\rho(t)\} \bigg) } &= \trace{ \rho(t) \bigg( L^\dagger_i O L_i - \frac{1}{2}\left\{L^\dagger_iL_i,O\right\} \bigg) }, 
\end{align}
so the expectation value evolves as
\begin{equation}
    \frac{\text{d}\braket{O}(t)}{\text{d}t} = \trace{ \rho(t) \left( i[H(t),O] + \sum_i \bigg( L^\dagger_i O L_i - \frac{1}{2}\left\{L^\dagger_iL_i,O\right\} \bigg) \right) }.
\end{equation}
In the Heisenberg picture, one usually defines $O_\text{H}(t)$ such that $\braket{O}(t) = \trace{O\rho(t)} = \trace{O_\text{H}(t)\rho(0)}$, thus the time evolution is absorbed in the observable itself instead of the state, picture that will be more convenient for our derivation, where the observable evolves following
\begin{equation}\label{eq:obs_open}
    \frac{\text{d}O_\text{H}(t)}{\text{d}t} = i[H(t), O_\text{H}(t)] + \sum_i \bigg( L^\dagger_iO_\text{H}(t)L_i - \frac{1}{2}\left\{L^\dagger_iL_i,O_\text{H}(t)\right\} \bigg),
\end{equation}
with the battery Hamiltonian our observable of interest. Recall that the battery Hamiltonian, after Jordan-Wigner transformation and Fourier transform, can be described in momentum space as $H_0=({g}/2)\sum_{i=1}^N \sigma^z_i = ({g}/2)\sum_{i=1}^N (1-2c^\dagger_i c_i) = ({g}/2)\sum_k (1-2c^\dagger_k c_k)$.

Let $n_k(t)= \braket{c^\dagger_kc_k}$ and $m_k(t)= \braket{c_{-k}c_k}$, with $\text{Re}[m_k(t)] = (\braket{c_{-k}c_k} + \braket{c^\dagger_kc^\dagger_{-k}})/2$ and $\text{Im}[m_k(t)] = (\braket{c_{-k}c_k} - \braket{c^\dagger_kc^\dagger_{-k}})/2i$. The coherent part of the evolution can be described as a system of $3N/2$ coupled linear ordinary differential equations, using the set of $3N/2$ variables $\{n_k(t), \text{Re}[m_k(t)], \text{Im}[m_k(t)]\}_{k>0}$, satisfying that $n_k(t)=n_{-k}(t)$ and $m_k(t)=m^\dagger_k(t)$. Therefore, the evolution of the energy stored can be computed as $E_N(t) = \trace{\rho(t)H_0} = ({g}/2) \big[ N - 4\sum_{k>0} n_k(t) \big]$.

With that, taking~\eqlabel{eq:kic_momentum} into account, the coherent part of~\eqlabel{eq:obs_open} reduces to
\begin{align}
    i[H(t), n_k(t)] &= -iJ\sin(ka) [c^\dagger_kc^\dagger_{-k} - c_{-k}c_k, c^\dagger_kc_k] = -iJ\sin(ka)( -c^\dagger_kc^\dagger_{-k} + c_{-k}c_k ) = \Delta_k\text{Im}[m_k(t)], \\
    \begin{split}
        i[H(t), \text{Re}[m_k(t)]] &= \frac{i}{2}[H(t), c_{-k}c_k + c^\dagger_kc^\dagger_{-k}] \\
        &= \frac{2iJ\cos(ka)}{2}[c^\dagger_kc_k, c_{-k}c_k + c^\dagger_kc^\dagger_{-k}] -\frac{iJ\sin(ka)}{2}\big( [c^\dagger_kc^\dagger_{-k}, c_{-k}c_k] + [c_{-k}c_k, c^\dagger_kc^\dagger_{-k}] \big) \\
        &= iJ\cos(ka)( -c_{-k}c_k + c^\dagger_kc^\dagger_{-k} ) = \epsilon_k \text{Im}[m_k(t)],
    \end{split} \\
    \begin{split}
        i[H(t), \text{Im}[m_k(t)]] &= \frac{i}{2i}[H(t), c_{-k}c_k - c^\dagger_kc^\dagger_{-k}] \\
        &= \frac{2J\cos(ka)}{2}[c^\dagger_kc_k, c_{-k}c_k - c^\dagger_kc^\dagger_{-k}] -\frac{J\sin(ka)}{2}\big( [c^\dagger_kc^\dagger_{-k}, c_{-k}c_k] - [c_{-k}c_k, c^\dagger_kc^\dagger_{-k}] \big) \\
        &= J\cos(ka)( -c_{-k}c_k - c^\dagger_kc^\dagger_{-k} ) - J\sin(ka)(1-2c^\dagger_kc_k) = -\epsilon_k \text{Re}[m_k(t)] +\frac{\Delta_k}{2}( 1 - 2n_k(t) ),
    \end{split}
\end{align}
where, recalling the free fermionic anticommutation formulas $\{c_i,c^\dagger_j\}=\delta_{ij}$ and $\{c_i,c_j\}=0$, we have used the following results:
\begin{align}
    [c^\dagger_kc^\dagger_{-k},c^\dagger_kc_k] &= c^\dagger_k(-c^\dagger_kc^\dagger_{-k})c_k - c^\dagger_k(1-c^\dagger_kc_k)c^\dagger_{-k} = -c^\dagger_kc^\dagger_{-k}, \\
    [c_{-k}c_k,c^\dagger_kc_k] &= c_{-k}c_k(1-c_kc^\dagger_k) - c^\dagger_k(-c_{-k}c_k)c_k = c_{-k}c_k, \\
    \begin{split}
    [c^\dagger_kc^\dagger_{-k},c_{-k}c_k] &= c^\dagger_k(1-c_{-k}c^\dagger_{-k})c_k - c_{-k}(1-c^\dagger_kc_k)c^\dagger_{-k} = c^\dagger_kc_k + c^\dagger_{-k}c_{-k} -1 = 2c^\dagger_kc_k -1,
    \end{split}
\end{align}
where in the last equality we use that $n_k(t)=n_{-k}(t)$.

Since $n_k(t)=n_{-k}(t)$ and $m_k(t)=m^\dagger_k(t)$ symmetries are preserved, it is possible to constraint the sum of~\eqlabel{eq:kic_momentum} to $k>0$ by simply substituting $\sum_k \mapsto 2\sum_{k>0}$, leading to the final system of $3N/2$ coupled linear ordinary differential equations
\begin{align}\label{eq:nk_coherent}
  \frac{\text{d}n_k(t)}{\text{d}t}\bigg|_\text{coh} &= 2\Delta_k\text{Im}[m_k(t)], \\
  \frac{\text{d}\text{Re}[m_k(t)]}{\text{d}t}\bigg|_\text{coh} &= 2\epsilon_k \text{Im}[m_k(t)], \\ \label{eq:immk_coherent}
  \frac{\text{d}\text{Im}[m_k(t)]}{\text{d}t}\bigg|_\text{coh} &= -2\epsilon_k \text{Re}[m_k(t)] + \Delta_k(1-2n_k(t)).
\end{align}

\subsection{Exact solution including dephasing}

After solving the coherent part of the evolution, in the following lines we extend its methodology to include different types of dissipative terms in the form of Lindblad operators, extending the system of coupled linear ordinary differential equations written in Eqs.~\eqref{eq:nk_coherent}-\eqref{eq:immk_coherent} with their corresponding dissipative contributions. In particular, we consider local-$\sigma^z$ dephasing and relaxation-pumping dissipation encoded as ladder operators.

In realistic implementations, coupling to an environment generically leads to decoherence. Local-$\sigma^z$ dephasing represents pure phase noise in the computational basis, inducing a decay of off-diagonal elements of the density matrix in the computational eigenbasis without involving spin flips, being a leading decoherence channel in many quantum platforms involving spin-chain Hamiltonians (related to $T_2$ decoherence)~\cite{fossfeig2013dynamical}. 

In our studies we consider local-$\sigma^z$ dephasing with a damping rate $\gamma_{z}$, namely $L_i=\sqrt{\gamma_{z}}\sigma^z_i$, returning
\begin{equation}\label{eq:dephasing}
  \frac{\text{d}\rho(t)}{\text{d}t} = -i[H(t), \rho(t)] + \gamma_{z}\sum_{i=1}^N \left( \sigma^z_i\rho(t)\sigma^z_i - \rho(t) \right).
\end{equation}
For this case, the time evolution is split into its continuous and kicked parts of the KIC Hamiltonian. The Lindblad master equation to be solved among subsequent kicks $t_{i-1}$ and $t_i$ is an autonomous equation that reads as
\begin{equation}\label{eq:cont_dephasing}
  \frac{\text{d}\rho(t)}{\text{d}t} = -i[H_I, \rho(t)] + \gamma_{z}\sum_{i=1}^N \left( \sigma^z_i\rho(t)\sigma^z_i - \rho(t) \right),\qquad (t_{i-1} < t < t_i)
\end{equation}
while the kick is applied as $\rho(t^+_i)= U_K\rho(t^-_i)U_K^\dagger$, with $U_K= e^{-i H_K}$ and $t^\pm_i$ the time $t_i$ right after or before the $i$th kick is applied, respectively. Therefore, between subsequent kicks, this methodology splits the evolution in a stepwise manner, starting from the Ising contribution through the master equation to evolve the density matrix for then applying the kick.

Rather than evolving analytically the density matrix, we can directly compute the energy injected dynamics under dephasing. In its Heisenberg form and for an observable $O(t)$,~\eqlabel{eq:cont_dephasing} transforms into
\begin{equation}\label{eq:heis_cont_dephasing}
  \frac{\text{d}O(t)}{\text{d}t} = i[H_I, O(t)] + \gamma_{z}\sum_{i=1}^N \left( \sigma^z_iO(t)\sigma^z_i - O(t) \right).
\end{equation}

While the coherent part of this evolution was derived in Eqs.~\eqref{eq:nk_coherent}-\eqref{eq:immk_coherent}, the dissipative contribution can be computed similarly. First, notice that after applying the Jordan-Wigner transformation, the dissipative part of~\eqlabel{eq:heis_cont_dephasing} simplifies into%
\begin{equation}\label{eq:diss_dephasing_fermion}
    \mathcal{D}[O(t)] = \sum_{j=1}^N \left( \sigma^z_jO(t)\sigma^z_j - O(t) \right) = \sum_{j=1}^N \left[ (1-2c^\dagger_jc_j)O(t)(1-2c^\dagger_jc_j) - O(t) \right] = -2\sum_{j=1}^N \left[ \{c^\dagger_jc_j,O(t)\}  - 2c^\dagger_jc_jO(t)c^\dagger_jc_j \right].
\end{equation}

At this stage,~\eqlabel{eq:diss_dephasing_fermion} is written under a free fermionic space but our variables are written using momentum space. To compute the dissipative contribution it is more convenient to apply the inverse Fourier transform of~\eqlabel{eq:fourier} to each variable, using that $c^\dagger_kc_k = (1/N)\sum_{m,n=1}^N e^{ik(m-n)}c^\dagger_mc_n$, $c_{-k}c_k = (i/N)\sum_{m,n=1}^N e^{-ik(m-n)}c_mc_n$ and $c^\dagger_kc^\dagger_{-k} = -(i/N)\sum_{m,n=1}^N e^{ik(m-n)}c^\dagger_mc^\dagger_n$, forms that we will use hereinafter.

We start by deriving the dissipative contribution corresponding to the number operator in momentum space $n_k=c^\dagger_kc_k$, where we have that
\begin{equation}\label{eq:nk_deph}
\begin{split}
    \mathcal{D}[c^\dagger_kc_k] &= -\frac{2}{N}\sum_{j,m,n=1}^N e^{ik(m-n)} \left( \{c^\dagger_jc_j, c^\dagger_mc_n\} -2c^\dagger_jc_jc^\dagger_mc_nc^\dagger_jc_j \right) \\
    &= -\frac{2}{N}\sum_{j,m,n=1}^N e^{ik(m-n)} \left(\delta_{jm}c^\dagger_jc_n + \delta_{jn}c^\dagger_mc_j -2\delta_{jm}\delta_{jn}c^\dagger_jc_j \right) \\
    &= -\frac{4}{N}\sum_{j,m=1}^N e^{ik(j-m)} c^\dagger_j c_m + \frac{4}{N} \sum_{j=1}^N c^\dagger_jc_j = -4\Bigg(c^\dagger_kc_k -\frac{1}{N}\sum_q c^\dagger_qc_q \Bigg),
\end{split}
\end{equation}
term that couples all momenta. For its calculation, we have used that
\begin{align}
    \begin{split}
    c^\dagger_jc_jc^\dagger_mc_n &= \delta_{jm} c^\dagger_jc_n - \delta_{jn} c^\dagger_mc_j + c^\dagger_mc_nc^\dagger_jc_j, 
    \end{split} \\ 
    \begin{split}
    c^\dagger_jc_jc^\dagger_mc_nc^\dagger_jc_j &= \delta_{jm}\delta_{jn} c^\dagger_jc_j - c^\dagger_m(\delta_{jn} - c_nc^\dagger_j)c_j = \delta_{jm}\delta_{jn} c^\dagger_jc_j -\delta_{jn}c^\dagger_mc_j + c^\dagger_mc_nc^\dagger_jc_j.
    \end{split}
\end{align}

Continuing with the real and imaginary parts of the anomalous operator $c_{-k}c_k$, terms that are present in the coherent part of the evolution [Eqs.~\eqref{eq:nk_coherent}-\eqref{eq:immk_coherent}], we obtain
\begin{align}\label{eq:mk_deph}
\begin{split}
    \mathcal{D}[c_{-k}c_k] &= \frac{2i}{N} \sum_{j,m,n=1}^N e^{-ik(m-n)} \left( \{c^\dagger_jc_j, c_mc_n\} -2c^\dagger_jc_jc_mc_nc^\dagger_jc_j \right) = \frac{2i}{N} \sum_{j,m,n=1}^N e^{-ik(m-n)} ( \delta_{jm}c_jc_n + \delta_{jn}c_mc_j ) \\
    &= \frac{4i}{N} \sum_{j,m=1}^N e^{-ik(j-m)} c_jc_m = -4c_{-k}c_k,
\end{split} \\ \label{eq:mk_dag_deph}
\begin{split}
    \mathcal{D}[c^\dagger_kc^\dagger_{-k}] &= -\frac{2i}{N} \sum_{j,m,n=1}^N e^{ik(m-n)} \left( \{c^\dagger_jc_j, c^\dagger_mc^\dagger_n\} -2c^\dagger_jc_jc^\dagger_mc^\dagger_nc^\dagger_jc_j \right) = -\frac{2i}{N} \sum_{j,m,n=1}^N e^{ik(m-n)} ( \delta_{jm}c^\dagger_jc^\dagger_n + \delta_{jn}c^\dagger_mc^\dagger_j ) \\
    &= -\frac{4i}{N} \sum_{j,m=1}^N e^{ik(j-m)} c^\dagger_jc^\dagger_m = -4c^\dagger_kc^\dagger_{-k},
\end{split}
\end{align}
implying that $\mathcal{D}[\text{Re}[m_k]] = -4\text{Re}[m_k]$ and $\mathcal{D}[\text{Im}[m_k]] = -4\text{Im}[m_k]$. For their derivation we have used that
\begin{align}
    \begin{split}
        c^\dagger_jc_jc_mc_n &= -\delta_{jm}c_jc_n -\delta_{jn}c_mc_j + c_mc_nc^\dagger_jc_j,
    \end{split} \\
    \begin{split}
        c^\dagger_jc_jc_mc_nc^\dagger_jc_j &= -\delta_{jm} c_jc_n -\delta_{jn}c_mc_j + c_mc_nc^\dagger_jc_j
    \end{split} \\
    \begin{split}
        c^\dagger_jc_jc^\dagger_mc^\dagger_n &= \delta_{jm}c^\dagger_jc^\dagger_n +\delta_{jn}c^\dagger_mc^\dagger_j + c^\dagger_mc^\dagger_nc^\dagger_jc_j,
    \end{split} \\
    \begin{split}
        c^\dagger_jc_jc^\dagger_mc^\dagger_nc^\dagger_jc_j &= c^\dagger_mc^\dagger_nc^\dagger_jc_j.
    \end{split}
\end{align}

It can be seen that, after Jordan-Wigner transform, Fourier transformation and some algebra;~\eqlabel{eq:heis_cont_dephasing} reduces to a system of $3N/2$ coupled linear ordinary differential equations
\begin{align}\label{eq:nk}
  \frac{\text{d}n_k(t)}{\text{d}t} &= 2\Delta_k\text{Im}[m_k(t)] -4\gamma_{z} \bigg( n_k(t) - \frac{2}{N}\sum_{q>0} n_q(t) \bigg) , \\
  \frac{\text{d}\text{Re}[m_k(t)]}{\text{d}t} &= 2\epsilon_k \text{Im}[m_k(t)] -4\gamma_{z} \text{Re}[m_k(t)], \\
  \frac{\text{d}\text{Im}[m_k(t)]}{\text{d}t} &= -2\epsilon_k \text{Re}[m_k(t)] + \Delta_k(1-2n_k(t)) -4\gamma_{z} \text{Im}[m_k(t)],
\end{align}
We can finally reconstruct the energy injected as $\braket{H_0} = ({g}/2)\sum_{i=1}^N \braket{\sigma^z_i}= ({g}/2)\big[N-4\sum_{k>0}\braket{n_k(t)}\big]$. 

While the aforementioned system solves the continuous part of the evolution under $H_I$ Hamiltonian, kicks can be applied using a $3\times 3$ decomposition per mode. Let $x_k(t)=\big[n_k(t), \text{Re}[m_k(t)], \text{Im}[m_k(t)]\big]^\text{T}$ and use that $U_Kc_qU^\dagger_K = e^{-2ib}c_q$ and $U_Kc^\dagger_qU^\dagger_K = e^{+2ib}c^\dagger_q$. Therefore, kicks can be applied as $x_k(t^+_i) = R^{(k)}_\text{kick}x_k(t^-_i)$, with the superindices of $t^{\mp}_i$ indicating that $x_k$ is evaluated at time $t_i$ befor and after applying the kick, respectively, with
\begin{equation}
  R_\text{kick} = \bigoplus_{k>0} R^{(k)}_\text{kick}, \qquad\text{with }R^{(k)}_\text{kick} = \begin{bmatrix}
    1 \\
    & \cos 4b & -\sin 4b \\
    & \sin 4b & \cos 4b
  \end{bmatrix}
\end{equation}
transformation that can be seen as a rotation of angle $4b$ over the $(\text{Re}[m_k(t)], \text{Im}[m_k(t)])$-plane. Consequently, the whole evolution is followed a stepwise procedure where the coherent evolution is solved, then the kick is applied, and the corresponding result is used to initialize the system for the subsequent kick.

\subsection{Thermal average of the energy stored in the battery}

In the charging dynamics analyses, the QB is initialized with a Gibbs state of the form
\begin{equation}\label{eq:initial}
    \rho(0) = \frac{e^{-\beta H_\text{th}}}{Z}, \qquad \text{with }Z=\trace{e^{-\beta H_\text{th}}}
\end{equation}
with $\beta=1/T$ the inverse temperature of the thermal bath, $H_\text{th}$ the Gibbs state Hamiltonian and $Z$ the partition function. We find two limit cases with respect $\beta$, where for the infinite-temperature limit we recover a completely mixed state, namely $\lim_{\beta\to 0} \rho(0) = 1/2^N$, and for the zero-temperature limit the initial state projects to the ground state of $H_\text{th}$, i.e., $\lim_{\beta\to\infty} \rho(0) = \proj{\text{GS}}$, since the Gibbs state acts as an imaginary-time propagator (with imaginary time $\tau=\beta$) and the long-time ($\tau\to\infty$) behaviour is dominated by its lowest eigenvalue~\cite{simon2005functional}. In our setup, the system evolves under the kicked-Ising model presented in~\eqlabel{eq:ref_ham1}, thus the state after $m$ kicks can be written as $\rho(m) = U(m)\rho(0)U^\dagger(m)$ with $U(m)=U^m(1)$ and $U(1) = \mathfrak{T}\exp\big[ -i\int_0^1 \text{d}t H(t)\big] = e^{-iH_K}e^{-iH_I}$.

To take finite-temperature effects into account, we consider the transverse field Ising model (TFIM) as Hamiltonian for the initial state [\eqlabel{eq:initial}], namely
\begin{equation}\label{eq:tfim}
    H_\text{th} = J_\text{th}\sum_{\braket{ij}} \sigma^x_i\sigma^x_j + h_\text{th}\sum_{i=1}^N \sigma^z_i.
\end{equation}
The same steps used in the previous section can be considered to diagonalize the TFIM model by decomposing it into $(k,-k)$ pairs, substituting $J\mapsto J_\text{th}$ and $b(t)\mapsto h_\text{th}$ in~\eqlabel{eq:kic_momentum}, which finally reads as 
\begin{equation}\label{eq:tfim_momentum}
\begin{split}
    H_\text{th} = \sum_k \left\{ 2[J_\text{th}\cos(ka) - h_\text{th}] c^\dagger_kc_k - J_\text{th}\sin(ka)(c^\dagger_kc^\dagger_{-k} + c_{-k}c_k) \right\} + \text{const.} = \sum_k \Psi^\dagger_k \mathcal{H}_{\text{th},k}\Psi_k + \text{const.}
\end{split}
\end{equation}
with $a$ the lattice spacing (which can be taken as $a=1$ for simplicity) and
\begin{equation}
  \mathcal{H}_{\text{th},k} = 2[J_\text{th}\cos(ka) - h_\text{th}]\tau^z - 2J_\text{th}\sin(ka)\tau^x = \epsilon_k\tau^z - \Delta_k\tau^x = \begin{bmatrix}
    \epsilon_k & -\Delta_k \\ -\Delta_k & -\epsilon_k
  \end{bmatrix}
\end{equation}
the Bogoliubov-de Gennes Hamiltonian per $k$ mode with eigenvalues $\lambda^\pm_k = \pm E_k = \pm\sqrt{\epsilon^2_k + \Delta^2_k}$, defining $\epsilon_k= 2[J_\text{th}\cos(ka) - h_\text{th}]$, $\Delta_k= 2J_\text{th}\sin(ka)$, and the Nambu spinor $\Psi_k = [c_k, c^\dagger_{-k}]^\text{T}$. However, to extract the exact charging dynamics, it is more convenient to span each $(k,-k)$ sector into a four-dimensional Hilbert space, namely $\Span{\ket{0_k0_{-k}}, \ket{1_k1_{-k}}, \ket{1_k0_{-k}}, \ket{0_k1_{-k}}}$. This decomposition yields $\braket{0_k0_{-k} | H_\text{th} | 0_k0_{-k}} = 0$, $\braket{1_k1_{-k} | H_\text{th} | 1_k1_{-k}} = 2\epsilon_k$, $\braket{0_k1_{-k} | H_\text{th} | 0_k1_{-k}} = \braket{1_k0_{-k} | H_\text{th} | 1_k0_{-k}} = \epsilon_k$ and $\braket{0_k0_{-k} | H_\text{th} | 1_k1_{-k}} = \braket{1_k1_{-k} | H_\text{th} | 0_k0_{-k}} = -\Delta_k$, returning the $4\times4$ Hamiltonian per $k$-mode
\begin{equation}
    H_{\text{th},k} = H^+_{\text{th},k} \oplus H^-_{\text{th},k} = \begin{bmatrix}
    0 & -\Delta_k \\
    -\Delta_k & 2\epsilon_k \\
    && \epsilon_k \\
    &&& \epsilon_k
  \end{bmatrix}, \qquad\text{with }H^+_{\text{th},k} = \epsilon_k(1-\tau^z) - \Delta_k\tau^x \text{ and }H^-_{\text{th},k} = \epsilon_k
\end{equation}
for $0<k<\pi$ and $H_{\text{th},k} = \epsilon_k(1-\tau^z)/2$ for the self-modes $k=0,\pi$. From this decomposition, we can clearly see that while the even sector $\Span{\ket{0_k0_{-k}}, \ket{1_k1_{-k}}}$ encapsulates the non-trivial dynamics, the odd sector presents a static non-zero contribution in general. In this four-dimensional subspace the battery Hamiltonian expands, after applying the Jordan-Wigner transformation and Fourier transform, as 
\begin{equation}\label{eq:batteryk}
  H_0 = \frac{{g}}{2}\sum_{i=1}^N \sigma^z_i = \frac{{g}}{2}\sum_{i=1}^N (1 - 2c^\dagger_ic_i) = \frac{{g}}{2}\sum_k( 1 - 2 c^\dagger_kc_k ) = \frac{{g}}{2} \sum_k O_k,
\end{equation}
with $O_k= 1-2n_k \equiv \diag(1,-1) = \tau^z$ if $k=0,\pi$ and $O_k=2(1 - n_k - n_{-k}) \equiv 2\diag(1,-1,0,0) = 2\tau^z\oplus 0_{2\times 2}$ if $0<k<\pi$, with $n_k= c^\dagger_kc_k$. Moreover, under this convention, the parity operator per $k$ mode becomes $P_k= (-1)^{n_k} = \tau^z$ if $k=0,\pi$ and $P_k= (-1)^{n_k + n_{-k}} = 1\oplus(-1)$ if $0<k<\pi$. 

Since $H_\text{th}$ is written as a sum of commuting Hamiltonians per mode, the inital Gibbs state factorizes as $\rho(0)=\bigotimes_k e^{-\beta H_{\text{th},k}}/Z_k$ with $Z_k=\trace{e^{-\beta H_{\text{th},k}}}$. Using the explicit $H_{\text{th},k}$ matricial forms, we have that
\begin{equation}
    Z_k = \begin{cases}
        1+e^{-\beta\epsilon_k} & \text{if }k=0,\pi \\
        2e^{-\beta\epsilon_k}[\cosh(\beta E_k)+ 1] & \text{if }0<k<\pi
    \end{cases}.
\end{equation}
As our analytical derivation is framed within momentum space, we have to work with states projected into a given fermionic parity sector, taking into account that the parity operator decomposes as $P=\prod_k P_k$. For this purpose, we will use that in real space, since $[H_\text{th},P]=0$, the Gibbs state can be rewritten as a sum of its projections onto the two sectors
\begin{equation}\label{eq:factor_rho}
    \rho(0) = \frac{P_+ H_\text{th}P_+ + P_- H_\text{th}P_-}{Z}, \qquad\text{with }Z=Z_+ + Z_-.
\end{equation}
The projected partition functions are obtained by restricting the fermionic traces to a fixed total fermion parity, which can be done by setting
\begin{equation}
    Z_\pm = \frac{1}{2}\left[ Z_{\{\text{AP},\text{P}\}} \pm Z^P_{\{\text{AP},\text{P}\}} \right], \qquad\text{with }Z_{\{\text{AP},\text{P}\}} = \prod_{k\in K_\pm} Z_k \quad\text{and}\quad Z^P_{\{\text{AP},\text{P}\}} = \prod_{k\in K_\pm} Z^P_k
\end{equation}
the parity-weighted single-mode traces for antiperiodic and periodic boundary conditions, respectively, with
\begin{equation}
    Z^P_k=\trace{P_ke^{-\beta H_{\text{th},k}}} = \begin{cases}
        1-e^{-\beta\epsilon_k} & \text{if }k=0,\pi \\
        2e^{-\beta\epsilon_k}[\cosh(\beta E_k) - 1] & \text{if }0<k<\pi
    \end{cases}.
\end{equation}
With that, we can easily compute the parity weights $w_\pm = Z_\pm/Z$ that will be employed later. Since $[H(t),P]=0$, the time evolution under the KIC Hamiltonian [\eqlabel{eq:ref_ham1}] will also preserve the parity, thus parity weights $w_\pm$ will be time-independent. Notably, the parity weights are also conserved along the evolution when pure dephasing is included [\eqlabel{eq:dephasing}], since $[H(t), P] = [L_i, P] = 0$, while for thermal dissipation [Eq.~(12) in the main text] no since $[\sigma^\pm_i, P] \neq 0$. For that case, it can be shown that the parity weights evolve as
\begin{equation}
    w_\pm(t) = \frac{1}{2}\big[ 1 \pm \braket{P}(t) \big] = \frac{1}{2}\left[ 1 + (2w_\pm(0)-1)e^{-N(\gamma_+ + \gamma_-)t} \right],
\end{equation}
which converge to $1/2$ at sufficiently large times.

Coming back to the thermal expectation derivation, to evolve on time each battery mode $O_k$ under the KIC Hamiltonian, we can compute their Heisenberg pictures on the even block using the $2\times2$ Floquet operator of~\eqlabel{eq:uk_floquet} as
\begin{equation}
    O^+_k(m) = 2U^\dagger_k(m)\tau^zU_k(m) = 2\left[ (1-2|\beta_k(m)|^2)\tau^z - 2\text{Re}[\alpha_k(m)\beta_k(m)]\tau^x - 2\text{Im}[\alpha_k(m)\beta_k(m)]\tau^y \right]
\end{equation}
for $0<k<\pi$, while the contribution of the self-modes $k=0,\pi$ remains constant along the evolution since $[\tau^z,U_{\{0,\pi\}}(m)]=0$. Therefore, using~\eqlabel{eq:factor_rho}, the thermal average of the energy stored in $H_0$ after $m$ kicks reduces to $E_N(m;\beta) = \braket{H_0(m)} = w_+ \braket{H_0(m)}_+ + w_- \braket{H_0(m)}_- $, with sector averages
\begin{equation}\label{eq:sector_average}
    \braket{H_0(m)}_\pm = \frac{g}{2}\sum_{k\in K_\pm} \frac{1}{2Z_\pm} \left[ A_k(m)Z_{\{\text{AP},\text{P}\}} \pm B_k(m)Z^P_{\{\text{AP},\text{P}\}} \right],
\end{equation}
and single-mode terms $A_k(m) = \trace{O_k(m)e^{-\beta H_{\text{th},k}}}/Z_k$ and $B_k(m) = \trace{O_k(m)P_ke^{-\beta H_{\text{th},k}}}/Z^P_k$, whose closed-form expressions can be computed as follows. For $0<k<\pi$, the even block thermal state reads as $\rho^+_k(0) = e^{-\beta H^+_{\text{th},k}}/\trace{e^{-\beta H^+_{\text{th},k}}} = \left[ 1 + \tanh\beta E_k(\epsilon_k\tau^z + \Delta_k\tau^x)/E_k \right]/2$. Tracing with the evolved operator $O^+_k(m)$, the $\tau^y$-term drops out giving
\begin{equation}
    \braket{O^+_k(m)} = 2\tanh\beta E_k \left[ \frac{\epsilon_k}{E_k}(1-2|\beta_k(m)|^2) + \frac{2\Delta_k}{E_k}\text{Re}[\alpha_k(m)\beta_k(m)] \right].
\end{equation}
Using that $Z^+_k=\trace{e^{-\beta H^+_{\text{th},k}}}=2e^{-\beta\epsilon_k}\cosh\beta E_k$ and that $Z^+_k/Z_k=\cosh\beta E_k/(\cosh \beta E_k + 1)$ and $Z^+_k/Z^P_k=\cosh\beta E_k/(\cosh \beta E_k - 1)$, we finally obtain that the single-mode terms are given by
\begin{align}
\begin{split}
    A_k(m) &= 2\tanh\frac{\beta E_k}{2} \left[ \frac{\epsilon_k}{E_k}(1-2|\beta_k(m)|^2) + \frac{2\Delta_k}{E_k}\text{Re}[\alpha_k(m)\beta_k(m)] \right],
\end{split} \\
\begin{split}
    B_k(m) &= 2\coth\frac{\beta E_k}{2} \left[ \frac{\epsilon_k}{E_k}(1-2|\beta_k(m)|^2) + \frac{2\Delta_k}{E_k}\text{Re}[\alpha_k(m)\beta_k(m)] \right],
\end{split}
\end{align}
where we have used the identities $\sinh 2x/(\cosh 2x \pm 1) = \{\tanh x, \coth x\}$. If $E_k=0$ ($J_\text{th} = h_\text{th} = 0$), then we would have that $\rho^+_k(0) = 1/2$ and $A_k=B_k=0$. For the self-modes $k=0,\pi$, the Hamiltonian is $H_{\text{th},k} = \epsilon_k(1-\tau^z)/2$ and $O_k=\tau^z$ is static, as aforementioned. Therefore, we obtain that $A_k=(1-e^{-\beta\epsilon_k})/Z_k$ and $B_k=Z_k/Z^P_k$. If $\epsilon_k=0$, we have that $Z^P_k=0$, so $B_k$ diverges. Nonetheless, the product $B_kZ^P_{\{\text{AP},\text{P}\}}$ in~\eqlabel{eq:sector_average} remains finite, so we can simply substitute this product by $Z_k\prod_{q\in K_\pm -\{k\}} Z^p_q$.

To conclude with this derivation, a special case can be found when $J_\text{th}$ is set to zero in~\eqlabel{eq:tfim}, thus the Gibbs state Hamiltonian leads to a local Hamiltonian, which can be factorized as $\rho_i = e^{-\beta h_\text{th}\sigma^z_i}/\trace{e^{-\beta h_\text{th}\sigma^z_i}}$, with $\rho(0) = \bigotimes_{i=1}^N \rho_i$. The exponential expands as $e^{-\beta h_\text{th}\sigma^z} = \cosh\beta h_\text{th} - \sinh\beta h_\text{th}\sigma^z \implies \trace{e^{-\beta h_\text{th}\sigma^z}} = 2\cosh\beta h_\text{th}$, returning that $\rho_i = (1-m^z_\beta\sigma^z_i)/2 = (1-\tanh\beta h_\text{th}\sigma^z_i)/2$, with $m^z_\beta= -\trace{e^{-\beta h_\text{th}\sigma^z}\sigma^z}/\trace{e^{-\beta h_\text{th}\sigma^z}}= \tanh\beta h_\text{th}$ the single-spin magnetization. For this particular case and defining $m^z_\beta= \tanh\beta h_\text{th}$, we obtain that 
\begin{equation}
    \braket{O_k(m)} = {g}m^z_\beta \bigg[ \sin^22J \sin^2k\frac{\sin^2 m\theta_k}{\sin^2\theta_k} - \frac{1}{2} \bigg],
\end{equation}
which further reduces at the self-dual point to $\braket{O_k(m)} = {g}m^z_\beta[ \sin^2mk - 1/2 ]$. Using~\eqlabel{eq:sector_average}, the energy stored in the battery Hamiltonian for this particular case is given by
\begin{equation}
    E_N(m;\beta) = \frac{gN}{2}\cdot\begin{cases}
        -m^z_\beta & \!\text{if }m\equiv 0 \!\!\!\pmod{N} \\
        (m^z_\beta)^{N-1} & \!\text{if }m\equiv \frac{N}{2} \!\!\!\pmod{N} \\
        0 & \!\text{otherwise}
    \end{cases},
\end{equation}
if $N$ is even, while for odd $N$ we obtain that
\begin{equation}
    E_N(m;\beta) = \frac{gN}{2}\cdot\begin{cases}
        -m^z_\beta & \!\text{if }m\equiv 0 \!\!\!\pmod{N} \\
        0 & \!\text{otherwise}
    \end{cases}.
\end{equation}

Defining the normalized energy injected as $\Delta E_N(m;\beta) = (E_N(m;\beta) - E_N(0;\beta))/N$, one finds that for even $N$, it is maximized at $m=N/2$, yielding $\Delta E_N(N/2;\beta) = ({g}/2) [m^z_\beta + (m^z_\beta)^{N-1}]$. This expression interpolates between zero and infinite temperature limits $\lim_{\beta\to\{0,\infty\}} \Delta E_N(N/2;\beta) = \{0,\text{sgn}(h_\text{th}){g}\}$, and, in the thermodynamic limit $\lim_{N\to\infty}\Delta E_N(N/2;\beta)=({g}/2)m^z_\beta$.%

\subsection{Concluding remarks: relationship between dissipative rates and \texorpdfstring{$T_1$}{T1} and \texorpdfstring{$T_2$}{T2} coherence times}

In the following lines, we establish a simple link between the dissipative rates associated with decoherence and thermal dissipation with respect to $T_2$ and $T_1$ coherence times, respectively, which can provide a measure of up to which extent the kicked-Ising QB remains functional before converging to a classical steady state. 

We begin with decoherence, introduced in our work via local Lindblad operators $L_i=\sqrt{\gamma_{z}}\sigma^z_i$, where we consider a single spin that evolves under under pure dephasing, namely
\begin{equation}\label{eq:1q_decoh}
    \frac{\text{d} \rho(t)}{\text{d}t} = \gamma_{z} ( \sigma^z \rho(t) \sigma^z - \rho(t) ) \implies \frac{\text{d}\phantom{t}}{\text{d}t} \begin{bmatrix} \rho_{00}(t) & \rho_{01}(t) \\ \rho_{10}(t) & \rho_{11}(t) \end{bmatrix} = -2\gamma_{z}\begin{bmatrix} 0 & \rho_{01}(t) \\ \rho_{10}(t) & 0 \end{bmatrix} \qquad\text{with }\rho(t) = \begin{bmatrix} \rho_{00}(t) & \rho_{01}(t) \\ \rho_{10}(t) & \rho_{11}(t) \end{bmatrix}.
\end{equation}
Therefore, we finally get that
\begin{align}
    \frac{\text{d}\rho_{\{00,11\}}(t)}{\text{d}t} &= 0 \implies \rho_{00}(t) = 1 - \rho_{11}(t) = \rho_{00}(0) = \text{const.}, \\
    \frac{\text{d}\rho_{\{01,10\}}(t)}{\text{d}t} &= -2\gamma_{z}\rho_{\{01,10\}}(t) \implies \rho_{\{01,10\}}(t) = \rho_{\{01,10\}}(0)e^{-2\gamma_{z} t}.
\end{align}

Defining $T_{\phi}$ as the timescale for which $2\gamma_{z} T_{\phi}\sim1$, we finally get that $T_{\phi} = 1/2\gamma_{z}$, which quantifies how coherences are lost over time due to dephasing purely. Notice that, since in our study we take as time interval between kicks $\tau=1$, there is a direct correspondence between the number of kicks applied and the continuous time $t$ under the evolution governed by~\eqlabel{eq:1q_decoh}.

We now turn to the the thermal dissipation contribution, introduced through the local Lindblad operators $L^\pm_i = \sqrt{\gamma_\pm}\sigma^\pm_i$ with $\sigma^\pm_i = (\sigma^x_i \pm i\sigma^y_i)/2$ the ladder operators, $\gamma_+=\gamma n_\text{th}$ and $\gamma_-=\gamma(n_\text{th}+1)$ the absorption and emission rates, respectively, with $n_\text{th}=1/(e^{\beta\omega_0}-1)$ the Bose-Einstein occupation number of a thermal bath at inverse temperature $\beta=1/T$ and frequency $\omega_0$. We consider again a single spin that evolves under the presence of thermal dissipation as
\begin{equation}\label{eq:1q_thdiss}
\begin{split}
    \frac{\text{d} \rho(t)}{\text{d}t} &= \gamma_+ \left( \sigma^+ \rho(t) \sigma^- - \frac{1}{2} \left\{ \sigma^-\sigma^+, \rho(t) \right\} \right) + \gamma_- \left( \sigma^- \rho(t) \sigma^+ - \frac{1}{2} \left\{ \sigma^+\sigma^-, \rho(t) \right\} \right) \\
    &\implies \frac{\text{d}\phantom{t}}{\text{d}t} \begin{bmatrix} \rho_{00}(t) & \rho_{01}(t) \\ \rho_{10}(t) & \rho_{11}(t) \end{bmatrix} = \begin{bmatrix} \gamma_+ - (\gamma_+ + \gamma_-)\rho_{00}(t) & -\frac{\gamma_+ + \gamma_-}{2}\rho_{01}(t) \\ -\frac{\gamma_+ + \gamma_-}{2}\rho_{10}(t) & -\left[\gamma_+ - (\gamma_+ + \gamma_-) \rho_{00}(t) \right] \end{bmatrix}, 
\end{split}
\end{equation}
where we have used that $\rho_{00}(t) + \rho_{11}(t) = 1$ in the last step. Therefore, we obtain that
\begin{align}\label{eq:th_diss1}
    \frac{\text{d}\rho_{\{01,10\}}(t)}{\text{d}t} &= -\frac{\gamma_+ + \gamma_-}{2} \rho_{\{01,10\}}(t) \implies \rho_{\{01,10\}}(t) = \rho_{\{01,10\}}(0)e^{-\frac{\gamma_+ + \gamma_-}{2} t}, \\ \label{eq:th_diss2}
    \frac{\text{d}\rho_{\{00,11\}}(t)}{\text{d}t} &= \pm \Big[ \gamma_+ - (\gamma_+ + \gamma_-)\rho_{00}(t) \Big] \implies \rho_{00}(t) = 1 - \rho_{11}(t) = \left( \rho_{00}(0) - \rho^\text{ss}_{00} \right)e^{-(\gamma_+ + \gamma_-) t} + \rho^\text{ss}_{00},
\end{align}
with $\rho^\text{ss}_{\{00, 11\}} = \lim_{t\to\infty} \rho_{\{00,11\}}(t) = \gamma_\pm/(\gamma_+ + \gamma_-) = [1 \mp \tanh( \beta\omega_0/2)]/2$ the excitation and ground-state steady state populations, respectively, with zero temperature limits $\lim_{\beta\to\infty} \rho^\text{ss}_{\{00, 11\}} = \{0,1\}$, as expected. Consequently, from Eqs.~\eqref{eq:th_diss1} and~\eqref{eq:th_diss2} we observe that thermal dissipation contributes to decohering off-diagonal terms on a timescale of $T_2$ as $(\gamma_+ + \gamma_-)T_2/2\sim1\implies T_2 = 2/(\gamma_+ + \gamma_-)$, and to the relaxation of the excited-state population on a timescale $T_1$ as $(\gamma_+ + \gamma_-)T_1\sim1\implies T_1 = 1/(\gamma_+ + \gamma_-) = T_2/2$. Using that $\gamma_+/\gamma_-=e^{-\beta\omega_0}$, we finally get that $1/T_1 = \gamma_-(1+e^{-\beta\omega_0})$, which is often rewritten by defining $\gamma_-=\Gamma_0/(1-e^{-\beta\omega_0})$ which finally leads to the universal thermal relation $1/T_1 = \Gamma_0\coth{\beta\omega_0/2}$.

When both local dephasing and thermal dissipation are taken into account, thus we sum the right-hand sides of Eqs.~\eqref{eq:1q_decoh} and~\eqref{eq:1q_thdiss}, we obtain that
\begin{equation}
    \frac{\text{d}\phantom{t}}{\text{d}t} \begin{bmatrix} \rho_{00}(t) & \rho_{01}(t) \\ \rho_{10}(t) & \rho_{11}(t) \end{bmatrix} = \begin{bmatrix} \gamma_+ - (\gamma_+ + \gamma_-)\rho_{00}(t)  & -\left( \frac{\gamma_+ + \gamma_-}{2} + 2\gamma_{z} \right)\rho_{01}(t) \\ -\left( \frac{\gamma_+ + \gamma_-}{2} + 2\gamma_{z} \right)\rho_{10}(t) & -\left[\gamma_+ - (\gamma_+ + \gamma_-)\rho_{00}(t) \right] \end{bmatrix},
\end{equation}
where now $T_2$ returns the standard relation
\begin{equation}
    \left( \frac{\gamma_+ + \gamma_-}{2} + 2\gamma_{z} \right)T_2\sim 1\implies \frac{1}{T_2} = \frac{\gamma_+ + \gamma_-}{2} + 2\gamma_{z} = \frac{1}{2T_1} + \frac{1}{T_{\phi}}.
\end{equation}

Finally, for both cases, physical units can be restored in terms of dimensionless units considering $\gamma_\text{phys} = \gamma/\tau_\text{phys}$, with $\tau_\text{phys}$ the period between kicks given in units of $1/|J|$ [\eqlabel{eq:ref_ham1}], with $J$ expressed as a frequency. Moreover, the physical temperature can be written in terms of dimensionless temperature $\beta=1/T$ as $T_\text{phys}/T=\hbar J_\text{th}/k_\text{B}$ with $J_\text{th}$ described as a frequency.%

\bibliography{bibfile}